**Grade inflation in introductory physics – the influence of out-of-class assignments**

Benjamin O. Tayo*

*Department of Physics, Pittsburg State University, Pittsburg KS 66762*

Ananda A. Jayawardhana

*Department of Mathematics, Pittsburg State University, Pittsburg, KS 66762*

**Abstract**

We report the results of statistical analysis performed on course grades for calculus-based introductory physics for data collected over a four-year period. We consider two important categories of scores: proctored (in-class proctored exams only) and proctored plus out-of-class (in-class proctored exams plus out-of-class assignments). The analysis revealed significant grade inflation in the proctored plus out-of-class scores. Quantile plots were used to compare the observed data and data modeled using the normal distribution. These plots revealed negligible correlation between the observed and modeled data for the proctored plus out-of-class scores, while a strong correlation is observed for the proctored scores. Using the proctored grade distribution as a reference, we performed goodness-of-fit tests using the Bayesian probability fit and the original reference proportions. Using the expected counts from the two different methods, we found p-values of 0.023 and 0.008. Both p-values support the hypothesis that there is significant difference in proctored plus out-of-class grade distribution compared to the reference. Further analysis showed that approximately 25% of all grades are shifted towards higher grades. Our studies clearly show that grade inflation induced by out-of-class assignments is a crucial issue in assessment that has to be addressed. By comparing the degree of inflation in our grade distribution with the national average, we found it to be about 50% less severe.

I. Introduction

Test scores play an important role in quantifying the intellectual ability of students. As educators, we need to ensure that the grade which a student receives at the end of a semester for a given course is well deserved and free from inflation. Most instructors use different methods to assess their students. In the physical sciences and for most classes taught at Pittsburg State University, in-class test (typically weighted at 50 to 80%) and out-of-class assignments (typically weighted at 20 to 50%) such as homework, quiz, project, and class attendance credit are often used as assessment methods. In an out-of-class assignment, students have all the time available at their disposal to get the assignment done. Sometimes they utilize resources such as the tutor room, google, and other resources to assist them in completing their assignments or they could simply copy a homework solution from their fellow classmates. Given that out-of-class assignments account for 20 - 50% of the overall grade which is a significant amount, we predict that with reference to the grades from proctored in-class exams, there is a significant inflation in the proctored plus out-of-class grade distribution. This hypothesis will be tested under the assumption that the data from the proctored exams can be treated as a good reference to the true grade distribution.

Grade inflation refers to the compression of grades towards higher grades [1]. The subject of grade inflation has been studied significantly over the past decade [1-2]. Grade inflation affects not only students, but also faculty members, administrators, as well as the reputation of the University. Studies have shown that when grades are highly inflated, it negatively impacts student's motivation to study hard [3], and students receiving inflated grades are not well-grounded in the fundamentals of the course material, which may lead to difficulties in upper-level courses [1]. Faculty may contribute to grade inflation by becoming more lenient which may cause them to assign high weights to out-of-class assignments in order to receive good evaluations from student, especially non-tenured faculties [4 - 6]. Administrators could also contribute towards grade inflation by promoting student retention efforts, which could drive faculties to compromise grading standards [7]. While several studies have been performed on the subject of grade inflation in the social sciences, the contribution to grade inflation from out-of-class assignments in an introductory-level physical science course has not been thoroughly investigated, to the best of our knowledge.

In this paper, we focus on the contribution to grade inflation from out-of-class assignments. We present the results of detailed statistical analysis performed on student data for introductory calculus-based physics taught at Pittsburg State University (Southeast Kansas). We compared the in-class proctored grades with the proctored plus out-of-class grades for data collected over a four-year period. We demonstrate using quantile plots that there is negligible correlation between the observed and modeled data for the proctored plus out-of-class scores, while a strong correlation is observed for the proctored scores. Using the proctored scores as a reference, we performed goodness-of-fit tests using the Bayesian probability fit and the original reference proportions. Using the expected counts from the two different methods, we found p-values of 0.023 and 0.008. Both p-values support the hypothesis that there is significant difference in proctored plus out-of-class grade distribution compared to the reference. Further analysis showed that approximately 25% of all grades are shifted towards higher grades. Our studies clearly show that grade inflation induced by out-of-class assignments is a crucial issue in assessment that has to be addressed. By comparing the degree of inflation in our grade distribution with the national average [8], we found it to be about 50% less severe.

These results will apply to most introductory level physical science courses and should guide educators in selecting appropriate weights for different assignment groups during syllabus preparation. It is recommended that the total weight assigned to out-of-class assignments be kept less than 15% in order to reduce the effects of grade inflation induced by out-of-class assignments. Studies have shown that by reducing the weights assigned to out-of-class assignments, and increasing the weight and number of in-class proctored exams, students become more motivated to learn and are able to retain the course material better [9].

This article is organized as follows. In section II, we described the details of data acquisition and general formalism. In section III, we discuss the results. A short summary concludes the report.

## II. Data Acquisition and General Formalism

Calculus-based introductory physics which is named engineering physics (PHYS 104) at Pittsburg State University is an introductory level course taken by science, engineering, and engineering technology students. The course covers the basics of mechanics, waves, fluids, and thermodynamics, and is offered every semester. The purpose of this work is to study and quantify the contribution to grade inflation from out-of-class assignments. To do so, we identify two important categories of scores, namely proctored and proctored plus out-of-class assignments. Proctored refers to the student scores from in-class proctored exams only, while proctored plus out-of-class refers to the student overall score at the end of the semester which includes contributions from in-class proctored exams and out-of-class assignments. On average, there is a total of 4 proctored tests per semester. Both proctored and proctored plus out-of-class scores are classified into 5 grade categories: A (90 – 100); B (80 – 89), C (70 – 79), D (60 – 69); and F (0 – 59).

In our studies, average proctored scores from 213 students and proctored plus out-of-class scores from 186 students were analyzed. For the proctored plus out-of-class scores, the total student count is 186, since we don't have overall scores for the 24 students currently enrolled in spring 2018, and a total of 3 students dropped out of the course during the past 7 semesters. These scores were collected for the following 8 semesters: fall 2014, spring 2015, fall 2015, spring 2016, fall 2016, spring 2017, fall 2017, and spring 2018. The students taking PHYS 104 typically are of the age group 18 to 21. For the 213 students studied, 86% were male and 14% were female. In terms of demographics, 87% were domestic and 13% were international. A majority of the domestic students are from the four-state region (Kansas, Missouri, Oklahoma, and Arkansas), while a majority of the international students are from Saudi Arabia, and a few from Asia and South America. These students were assessed based on in-class proctored exams and out-of-class assignments (homework, quiz, project, and class attendance). For the 8 semesters considered, in-class proctored tests constitute on average 63% of the overall grade, while out-of-class assignments account for 37% of the overall grade.

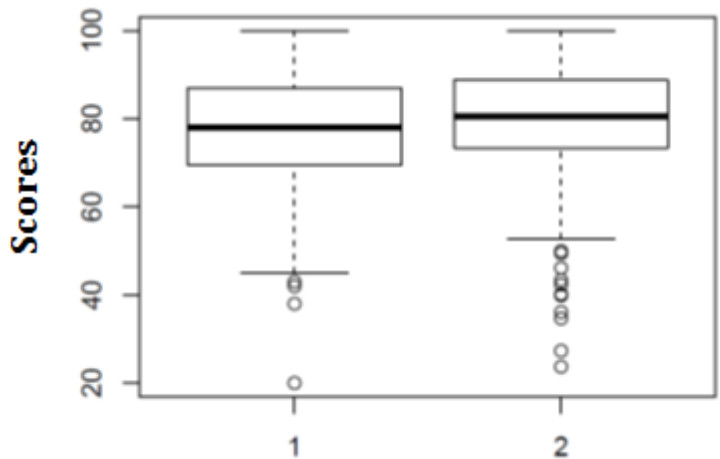

Figure 1. Boxplot for proctored (1) and proctored plus out-of-class (2) scores. The proctored scores are symmetric around the mean, as expected for the normal distribution, while the proctored plus out-of-class scores are not.

The student data for the past 8 semesters was exported from the course management website (canvas) and saved in different Excel files. R Sudio was used for data analysis. A code was written to import student scores and convert the scores into the appropriate grade category. The function *boxplot* was used to display the range and average for the data. The average score and standard deviation were calculated using the functions *mean* and *sd*, respectively. The function *prop.table* was used to calculate the observed grade proportions. The function *quantile* was used to calculate the observed percentiles for the scores. The function *qnorm* was used to calculate the expected percentiles for both the proctored and proctored plus out-of-class scores. The observed percentiles were plotted against the expected percentiles for both the proctored and proctored plus out-of-class grades in order to determine if grades are distributed according to the normal distribution. The function *abline* was used to add the identical line for the observed versus expected quantile plots. Finally, a Chi-Square goodness-of-fit test was performed using Manitab in order to determine the degree of correlation between the proctored and proctored plus out-of-class grades.

III. Results and Discussions

All calculations were performed using the following softwares: R Studio, EXCEL, WinBUGS, and Manitab. In Fig. 1, we show the boxplot for the proctored and proctored plus out-of-class scores. Except for a few outliers, the proctored scores have a range of 45.0 to 100.0, with a median of 78.0. The 25[th] and 75[th] percentile scores being 69.5 and 87.0, respectively. For the proctored plus out-of-class scores, the range is 52.8 to 100.0, with a median of 80.6. The 25[th] and 75[th] percentiles scores are 73.4 and 88.8, respectively. The proctored score distribution is approximately symmetric, indicating that proctored scores are normally distributed, while the proctored plus out-of-class scores are non-symmetric, and hence do not obey the normal distribution. The proctored scores have a mean of 77.5 and a standard deviation of 13.6, while the proctored plus out-of-class scores have a mean of 78.9, and a standard deviation of 14.5. The median of the proctored score is 78.0, which is approximately equal to the mean (77.5), as expected for a symmetric normal distribution. The median of the proctored plus out-of-class scores is 80.6, which is greater than the mean (78.9), indicating a non-symmetric distribution.

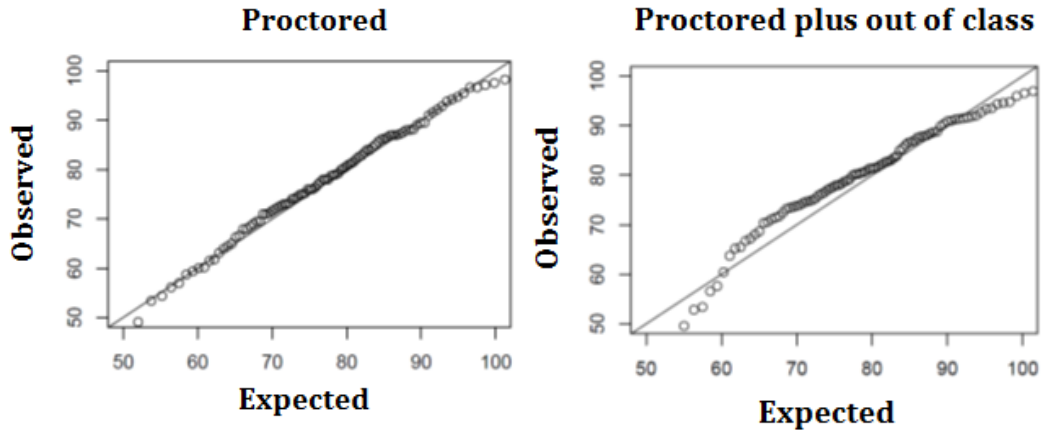

Figure 2. Observed versus predicted quantile plots for proctored and proctored plus out-of-class work scores. The solid line indicates the identical line.

To further confirm that the proctored scores follow the normal distribution, while the proctored plus out-of-class scores do not, we plotted the observed versus expected percentiles for both the proctored and proctored plus out-of-class scores. Figure 2 shows the observed percentiles plotted against the expected percentiles calculated using the normal distribution. The solid line on both plots indicate the identical line. Figure 2 shows clearly that the proctored exam scores are distributed according to the normal distribution as most of the data points fit nicely with the identical line. However, for the proctored plus out-of-class scores, the expected percentiles do not correlate well with the observed percentiles. Hence we conclude that the proctored plus out-of-class scores are significantly biased due to the contribution from out-of-class assignments.

Following the analysis above, we observe that the proctored scores are distributed according to the normal distribution and hence can be treated as a good reference for the true grade distribution. Under this assumption, we performed a Chi-Square goodness-of-fit test using Manitab in order to determine the degree of correlation between the proctored scores and the proctored plus out-of-class work scores, using the proctored scores as the reference or baseline.

Reference proportions were found in two different ways: 1) Using data fitted as a multinomial distribution and using Bayesian methods and WinBUGS software with uniform prior distributions selected with limits as instructor's knowledge of percentage limits of each grade (see Table 1); and 2) Using the simple fraction of each grade class (see Table 2). Multiplying these estimates by the total of 186 students, we found the expected counts for the proctored plus-out-of-class grades. Figure 3 shows the Chi-Square goodness-of-fit test for observed and expected grade counts, with the expected values calculated using reference proportions obtained with the Bayesian multinomial probability fit. Figure 4 shows the Chi-Square Goodness-of-Fit test for observed and expected grade counts, with the expected values calculated using the simple reference proportions. Using the expected counts from the two different methods, we found p-values of 0.023 and 0.008. Both p-values support the hypothesis that there is significant difference in the proctored plus out-of-class grade distribution compared to the reference. Since the individual data show that there are more than expected A and B grade, about the same C grades, less than expected D grades, and about the same F grades, we conclude that the difference is towards grade inflation.

Table 1. Observed Counts and expected counts calculated using the Bayesian multinomial probability fit.

| Grade | Proctored (Reference) | | Proctored plus out-of-class | |
|---|---|---|---|---|
| | Observed Counts | Bayesian Multinomial Probability Fit | Observed Counts | Expected Counts Using the Reference Proportions |
| A | 36 | 0.183 | 45 | 34.0 |
| B | 60 | 0.274 | 55 | 50.9 |
| C | 63 | 0.291 | 53 | 54.2 |
| D | 33 | 0.154 | 14 | 28.6 |
| F | 21 | 0.098 | 19 | 18.3 |
| Total | 213 | 1.000 | 186 | 186 |

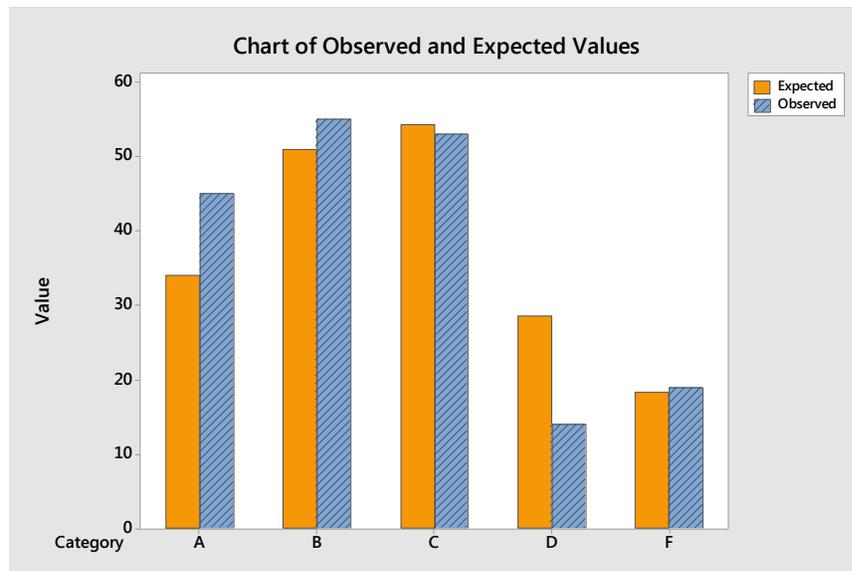

Figure 3. Chi-Square goodness-of-fit test for observed counts in variable: count. Expected values calculated using reference proportions obtained with the Bayesian multinomial probability fit.

Comparing the observed proctored plus out-of-class grade proportions with the reference proctored grade proportions in Table 2, we also observe that the proportion of students receiving an A grade from proctored exams is 16.9% while that from proctored plus out-of-class assignments is 24.2%. The additional 7.3% increase reflects the contribution from out-of-class assignments. The proportion of students receiving a B, C or F grade remains approximately the same for both the proctored and the proctored plus out-of-class grades. For the D grade, the proportion reduces from 15.5% for proctored grades to 7.5% for proctored plus out-of-class grades, which is an 8 % reduction. This means considering the biases introduced by out-of-class assignments, 8% of D students become C students, and 8% of C students become B students, while 8% of B students become A students. The net effect is 8% more A students, and 8% less D students. Thus, considering the grade transitions from D to C, C to B, and B to A, we notice that approximately 24% of all grades are shifted towards higher grades.

Table 2: Observed counts and expected counts calculated using simple proportion of grades from reference grade distribution.

| Grade | Proctored Exams (Reference) | | Proctored plus out-of-class | |
|---|---|---|---|---|
| | Observed Counts | Proportion =Count/Total | Observed Counts | Expected Counts Using the Reference Proportions |
| A | 36 | 0.169 | 45 | 31.4 |
| B | 60 | 0.282 | 55 | 52.4 |
| C | 63 | 0.296 | 53 | 55.0 |
| D | 33 | 0.155 | 14 | 28.8 |
| F | 21 | 0.098 | 19 | 18.4 |
| Total | 213 | 1.000 | 186 | 186 |

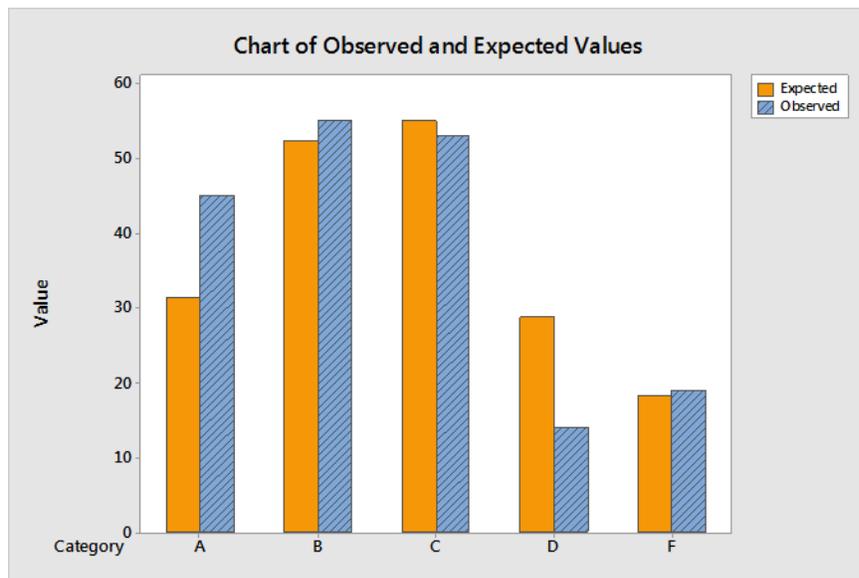

Figure 4. Chi-Square goodness-of-fit test for Observed Counts in Variable: Count. Expected values calculated using simple proportions from the reference grade distributions

From Table 1, we remark that even with a 37% weighting contribution from out-of-class assignment to the overall grade, letter grade A is not the most frequent grade, accounting for 24% of all letter grades. We also observe that D's and F's in total account for about 18% of all letter grades. A study conducted across a wide range of institutions revealed that 43% of all grades were A, while D's and F's accounted for less than 10% of all grades [8]. Comparing these with our studies, we remark that grade inflation in calculus-based introductory physics at Pittsburg State University is approximately 50% less severe, when compared to other institutions.

## IV. Conclusions

In summary, we have shown using statistical analysis that out-of-class assignments when weighted at 37% of final score, do contribute significantly towards grade inflation. By computing and comparing the observed and expected percentiles for proctored and proctored plus out-of-class scores, we found that the proctored scores are correctly distributed according to the normal distribution while the proctored plus out-of-class scores are not. Using the proctored scores as a reference, we performed goodness-of-fit tests using the Bayesian probability fit and the original reference proportions. Using the expected counts from the two different methods, we found p-values of 0.023 and 0.008. Both p-values support the hypothesis that there is significant difference in proctored plus-out-of-class assignments compared to the reference. Further analysis showed that 25% of all grades are shifted towards higher grades. Our studies clearly show that grade inflation induced by out-of-class assignments is a crucial issue in assessment that has to be addressed. By comparing the degree of inflation in our grade distribution with the national average, we found it to be about 50% less severe. We recommend reducing the weight assigned to out-of-class assignments to less than 15% and increasing the weight assigned to in-class exams as a means to reduce grade inflation.

**Acknowledgement**

The authors acknowledge useful discussions with Dr. Janet S. Zepernick from the department of English and Modern Languages at Pittsburg State University.